\documentclass[11pt]{article} 

\usepackage[utf8]{inputenc} 

\usepackage{authblk}

\usepackage{geometry} 
\usepackage{graphicx} 

\usepackage{booktabs} 
\usepackage{array} 
\usepackage{paralist} 
\usepackage{verbatim} 
\usepackage{subfig} 
\usepackage{amssymb, amsmath, amsthm, mathabx}
\usepackage{epstopdf}

\usepackage{algorithm2e}

\usepackage{fancyhdr} 
\usepackage{sectsty}
\allsectionsfont{\sffamily\mdseries\upshape} 

\usepackage[nottoc,notlof,notlot]{tocbibind} 
\usepackage[titles,subfigure]{tocloft}
\usepackage{mathrsfs}
\usepackage{bm}
\usepackage{color}
\usepackage{bbm}

\usepackage{cite}
\usepackage{amsmath,amssymb,amsfonts,graphicx}
\usepackage{bm}
\usepackage[nottoc,notlof,notlot]{tocbibind}
\usepackage{animate}

\newcommand{\R}{\mathbb{R}}
\renewcommand{\S}{\mathbb{S}}
\newcommand{\dd}{\mathrm{ d}}
\newcommand{\PP}{\textbf{Gr}(1,3)}

\newtheorem{theorem}{Theorem}[section]

\newtheorem{conjecture*}{Conjecture}

\graphicspath{{./figs/}}

\title{John Transform and Ultrahyperbolic Equation for Lightfields}
\author{Todor Georgiev, He Qin, Haotian Li}
\affil{Adobe Systems, San Jose, CA 95110, USA}
\date{\today}

\begin{document}

\maketitle
\section*{Abstract}
This paper explores possibilities for new uses of the Radon transform for imaging and analysis of lightfields. We show that the previously reported Dimansionality Gap \cite{5539854} can be derived from an ultrahyperbolic PDE, first proposed by F. John \cite{john1985ultrahyperbolic}, which is satisfied by lightfields. 
Based on inverse John transform we demonstrate rigorous Focal Stack rendering and viewing from arbitrary angles. Based on Asgeirsson's theorems for the ultrahyperbolic PDE we derive new kernels for processing lightfields. Our kernels provide alternative methods for depth computation and other image processing in lightfields.

\noindent \textbf{Keywords}: Radon transform, ultrahyperbolic differential equation, John's equation, $4D$ Radiance, Kernel, lightfield, depth estimation.

\section{Introduction}
The Radon transform of a function $f$ in $n$-dimensional space is defined as the integrals of $f$ over all $k$-dimensional planes, where $k$ is fixed and $k<n$ \cite{gelfand1966generalized,helgason1999radon}. There are $n-1$ Radon transforms, each parametrized by the parameters that are required to describe the planes. The manifold of all $k$-planes in $n$-dimensional space is known as the Grassmann manifold, and it is denoted by $\textbf{Gr}(k,n)$. Dimensionality is defined by the number of parameters of the transform. For example, the specific transform for $k=n-1$ depends on $n$ parameters.

\begin{figure}[hbt!]
    \centering
    \includegraphics[scale = 0.3]{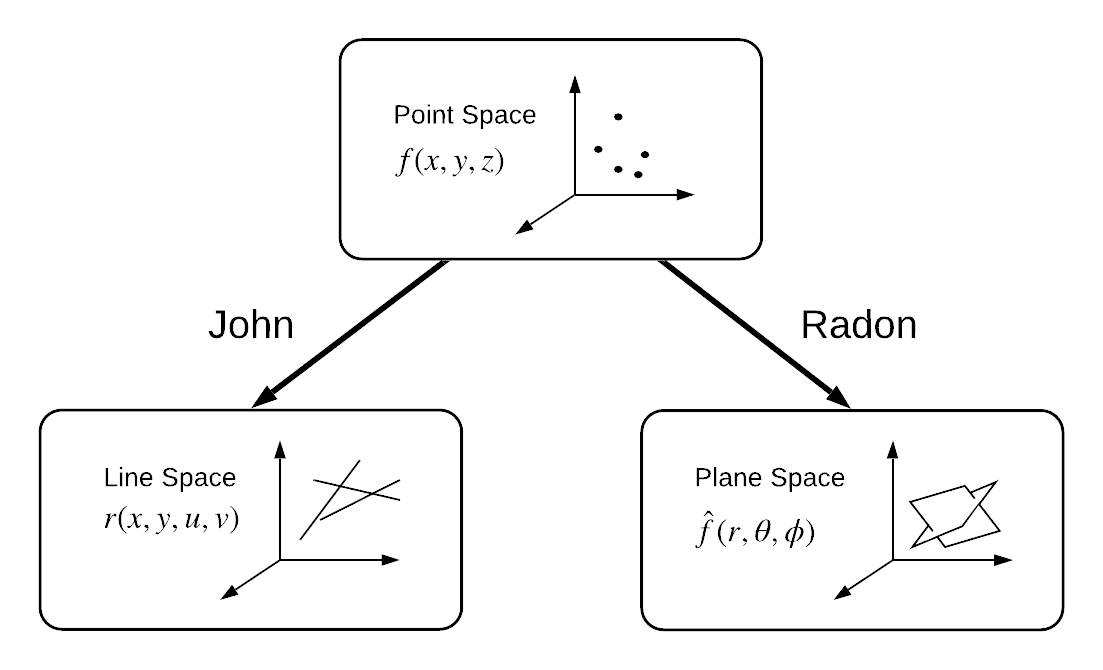}
    \caption{Radon and John Transform}
    \label{fig: RadonJohnRelationship}
\end{figure}

In $3D$ there are two different Radon transforms, respectively associated with lines and planes, as shown in Fig.\ref{fig: RadonJohnRelationship}. The manifold of all lines in $3D$ is $\PP$, which stands for ``all $1D$ planes in $3D$.'' As we know, it is  $4$-dimensional: That's the lightfield. The space of all planes is  $\textbf{Gr}(2,3)$, i.e., ``all $2D$ planes in $3D$''. It is $3$-dimensional. In this paper we are dealing with the Radon transform associated with lines, i.e., the mapping of $f(x, y, z)$ into functions in $\PP$. This mapping is one of the Radon transforms, but it also has its own name. It is called ``the John Transform'' after a seminal paper \cite{john1985ultrahyperbolic} that first introduced it and investigated its properties for X-ray imaging.  

The Radon transform can be inverted and there are well studied formulas and algorithms for inversion in each case\cite{buzug2012computed, deans1983radon, macovski1983medical}. Most of the applications come from the ability to invert the transform.

We will consider the following two applications that are of interest for computational photography. (1) Under the Lambertian condition, acquisition of focal stack images is by its nature the John transform. (2) Lightfield imaging can also be depicted using John's transform, and the lightfield generally satisfies John's equation.

Considering the first application, we show how the focal stack can be inverted to produce the distribution of light sources $f(x, y, z)$ in $3D$ space. This is the inverse John transform. Then we demonstrate that viewing $f$ from different angles using John's transform produces realistic renderings of full 3D parallax in a very wide range of angles. This broadens the area where effects  like \cite{5539854} can be implemented.

Considering the second application, $\textbf{Gr}(2,3)$ is $3$-dimensional. It has the same dimensionality as the original $3D$ space. In contrast, $\PP$ is $4$-dimensional. That's the lightfield space describing the radiance. In general, if radiance is generated from some substance distributed in $3D$ and having isotropic scattering or absorption, then radiance could not be $4$-dimensional: It is generated by a function of 3 parameters. Since by its nature radiance depends on 4 parameters, it follows that there must be one constraint. Our paper shows that this is the root of the so called Dimensionality Gap, i.e., the observation that even though the lightfield is $4$-dimensional, in the Fourier transform it is nonzero only on a special $3D$ manifold \cite{5539854, Levin:2009:FAC:1576246.1531403, Ng:2005:FSP:1073204.1073256}.

We derive from first principles the partial differential equation satisfied by the radiance, which equation defines the above constraint. It turns out to be a form of the Ultyrahyperbolic PDE. The solutions of this equation satisfy two theorems by Asgeirsson (see \cite{courant2008methods}). We derive the discretized form of some of the above Asgeirsson's relations. These are new results which can be used for finding depth, for denoising, or other image processing purposes. Filtering based on Asgeirsson's theorems is not equivalent to previous methods. This paper will also show some of our experimental results.

\section{The John Transform}
\subsection{Forward Transform}
\begin{figure}[hbt!]
    \centering
    \includegraphics[scale = 0.4]{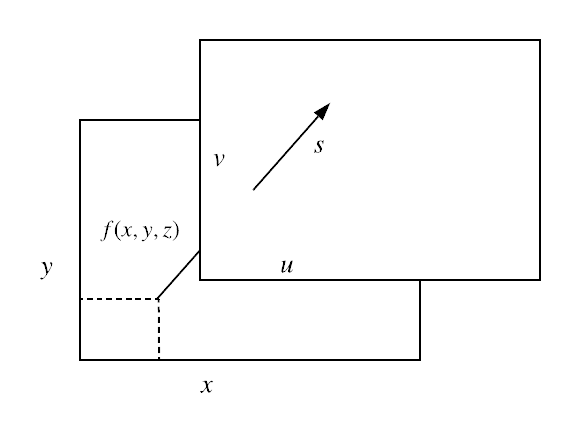}
    \caption{Two parallel planes parameterization of straight lines in $3D$.}
    \label{fig: JT}
\end{figure}

Given a function $f$ describing density of isotropic light sources in $3D$, the John transform $r$ of $f$ is defined as its integral along a straight line:
\begin{align}
    J(f) = r, \qquad r(\xi) := \int_{\xi} f(x,y,z) \dd m(x,y,z), \quad \xi \in \PP
\end{align}
where $\dd m$ is the Euclidean measure on the straight line $\xi$. If we use two parallel planes parametrization for $\xi$ (Fig.\ref{fig: JT}), where $(x,y)$ gives the intersection of light ray $\xi$ with the horizontal plane and $(u,v)$ indicates the angles by tracking the displacements of $\xi$ on the second plane, then
\begin{align}
    J(f) = r, \qquad r(x,y,u,v) = \int_{-\infty}^\infty f(x + u z, y + v z, z) dz
    \label{eqn: John Transform}
\end{align}
The radiance $r$ is the quantity captured by the plenoptic camera. Conventional cameras capture the integral of $r$ over the aperture. Other cameras also capture certain integrals of the radiance.

\subsection{Inverse John Transform}
In order to get back the density $f$ of light sources based on the knowledge of the radiance $r$, we need two main steps, i.e., (1) computing the dual transform and (2) deconvolution from the dual transform by the appropriate kernel. We base our discussion on \cite{helgason1999radon}.

First, we give the form definition of the dual transform.
The dual transform $\check{r}$ is defined, for each point $p = (x,y,z)$ in the $3D$ point space, as the integral average of the radiance $r$ over all straight lines through $p$. We use the notation $\check{p} \subset \PP$ to indicate the pencil of lines through point $p$.
\begin{align}
    \check{p} = \{ \xi_{p,w} \in \PP | \xi_{p,w}: p + s\cdot w = (x+sw_1,y+sw_2,z+sw_3), s\in \R \}
    \label{eqn: direction parameterize}
\end{align}
where $w = (w_1,w_2,w_3) \in \S^2$ is the unit direction vector for $\xi_{p,w} \in \check{p}$ and note that $\xi_{p,w} = \xi_{p,-w}$ (undirected lines). Also, note that since the area of the unit sphere is $4\pi$, the dual transform is given by
\begin{align}
    \Check{r}(x,y,z) := \frac{1}{4\pi}\int_{\S^2} r (\xi_{p,w}) \dd w 
    \label{eqn: dual transform}
\end{align}

If we substitute $r$ with the John transform Eq.\ref{eqn: John Transform}, then we can write Eq.\ref{eqn: dual transform} in the following way (Chapter I, \cite{helgason1999radon}).
\begin{align}
    \Check{r}(x,y,z) 
     = \frac{1}{4\pi}   \int_{\S^{2}} \int_{-\infty}^\infty f(x + s w_1, y+s w_2, z+s w_3 ) \dd s \dd w 
\end{align}
By change of variables, $x' = x + sw_1$, $y' = y + s w_2$, $z' = z + sw_3$, and $\dd x' \dd y' \dd z' = s^2 \dd s \dd w$,
\begin{align}
    \Check{r}(x,y,z) 
     = \frac{1}{2\pi} \int_{\R^3} \frac{1}{(x-x')^2 + (y-y')^2 + (z-z')^2} f(x', y', z') \dd x' \dd y' \dd z' = \frac{1}{2\pi} (D^{-2} * f)
     \label{eqn: r check}
\end{align}
where $D = D(x,y,z) := \sqrt{x^2+y^2+z^2}$ and $*$ is the $3D$ convolution.

Now for the second step of the inverse John transform, we deconvolve Eq.\ref{eqn: r check} to get the original light source density $f$. This can be done in the frequency domain. From Lemma 5.2 in Chapter V of \cite{helgason1999radon}, we know that the Fourier transform of $D^{-2}$ is $\mathcal{F}(D^{-2}) = (D^{-2})^\sim = 2\pi^2 D^{-1}$. Therefore,
\begin{align*}
    (\Check{r})^{\sim} = \frac{1}{2\pi} (D^{-2})^\sim \cdot \Tilde{f} &=  \frac{1}{2\pi} (2\pi^2 D^{-1}) \cdot \Tilde{f} = \pi D^{-1} \cdot \Tilde{f} 
\end{align*}
\begin{align}
    \implies \Tilde{f} &= \frac{1}{\pi} D \cdot (\Check{r})^{\sim}
    \label{eqn: Fourier inverse John}
\end{align}
Then, one can get $f$ by the inverse Fourier transform. 
In Helgason's book (Chapter V, \cite{helgason1999radon}), he introduces the fractional powers of the Laplacian which can be used to describe Eq.\ref{eqn: Fourier inverse John} in the original spatial domain.
\begin{align}
    f = \frac{1}{\pi} \sqrt{-\Delta} \text{ } \Check{r}
    \label{eqn: inverse John transform}
\end{align}
But essentially, Eq.\ref{eqn: Fourier inverse John} and Eq.\ref{eqn: inverse John transform} are the same under Fourier transform.

The immediate application of Eq.\ref{eqn: Fourier inverse John} is to render a $3D$ scene from focal stack.

\subsection{3D from Focal Stack}

\begin{figure}[hbt!]
    \centering
    \includegraphics[scale = 0.35]{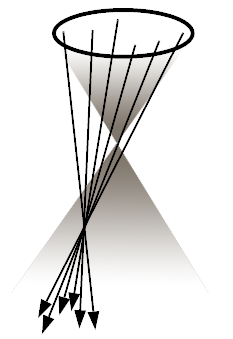}
    \caption{All rays of the focusing cone at a pixel. Another focusing cone, represented as a shadow, adds more energy to that pixel.}
    \label{fig: dual transform}
\end{figure}

Data capture into a focal stack can be understood in the following way.
At any point in the physical focal stack, the recorded light intensity is the sum of the energy of all rays of the focusing cone at that particular pixel plus the energy of any other rays that happen to reach that same pixel without being focused on it (see Fig.\ref{fig: dual transform}). This sum can be approximated by the dual John transform $\check{r}$ at this point.

\begin{figure}[hbt!]
    \centering
    \includegraphics[scale = 0.5]{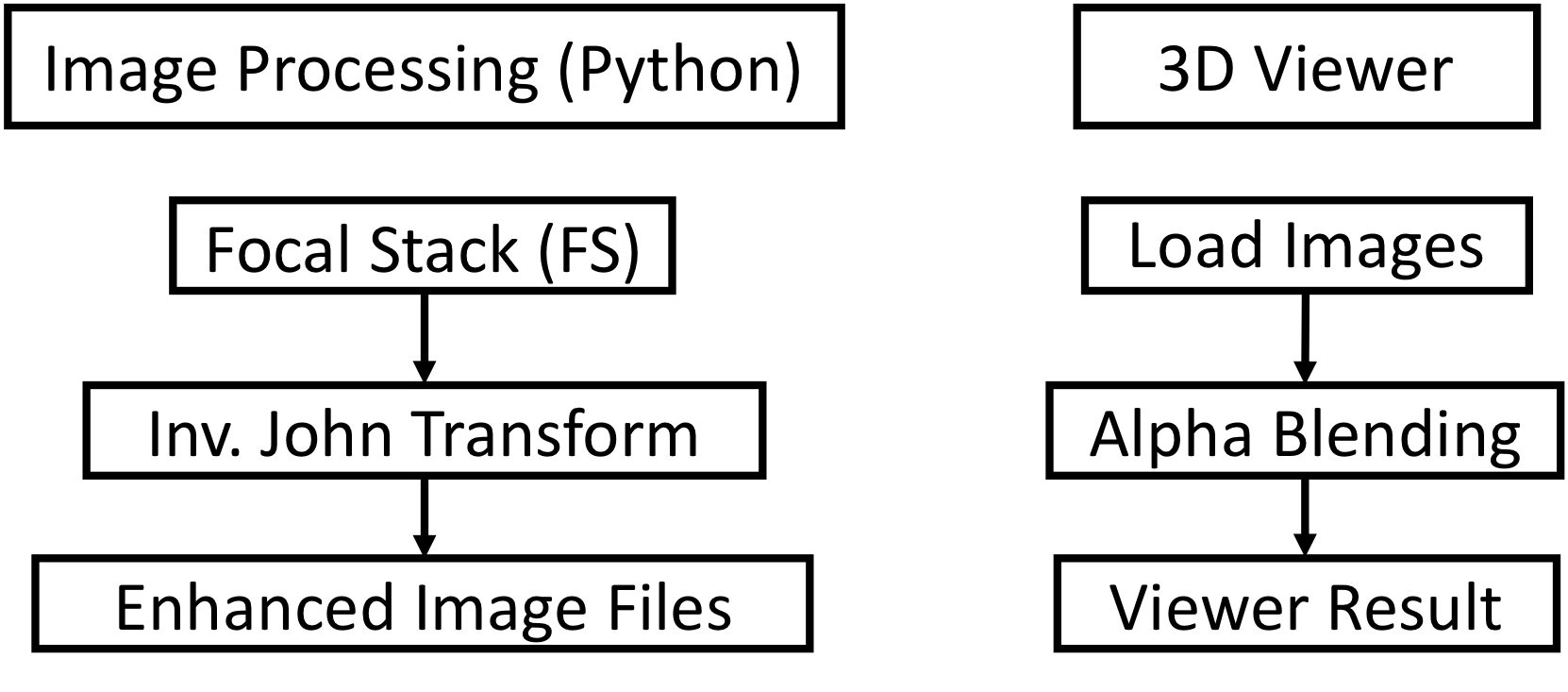}
    \caption{Pipeline for Inverse Radon Transform on Focal Stack}
    \label{fig: focalstackExperiment}
\end{figure}

The inverse John transform renders the $3D$ scene from a set of $2D$ images captured in a focal stack. In our experiment, the input focal stack is $13$ images of an ant taken from \cite{Photomontage}. This experiment consists of two parts. Firstly, we pre-process all images of the focal stack by applying inverse John transform. Secondly, we visualize the $3D$ result implementing the John transform over virtual camera rays integrating intensity with equal Alpha blending. The block diagram of the experiment is shown in Fig.\ref{fig: focalstackExperiment}.
According to the previous section, the inverse John transform in $3D$ can be implemented using Fourier transform. The images are first loaded into a 3-dimensional array. After $3D$ Fourier transform of the array, we multiply it by (Fourier transformed) square root of the Laplacian, i.e., multiply it by $D$ (see Eq.\ref{eqn: Fourier inverse John}). Then we perform the inverse Fourier transform. Notice that this operator changes brightness. We can correct brightness using the Parseval's theorem,
\begin{align}
    \int_{-\infty}^{\infty}|f(x)|^2 \dd x = \int_{-\infty}^{\infty}|\tilde{f}(k)|^2 \dd k.
    \label{eqn: Parseval theorem}
\end{align}
The resulting color is still dim and blurry, so we linearly scale it with Levels, 5/95 percentile. With this, the transformed images are ready for viewing.

\begin{figure}[hbt!]
    \centering
    \includegraphics[scale = 1.0]{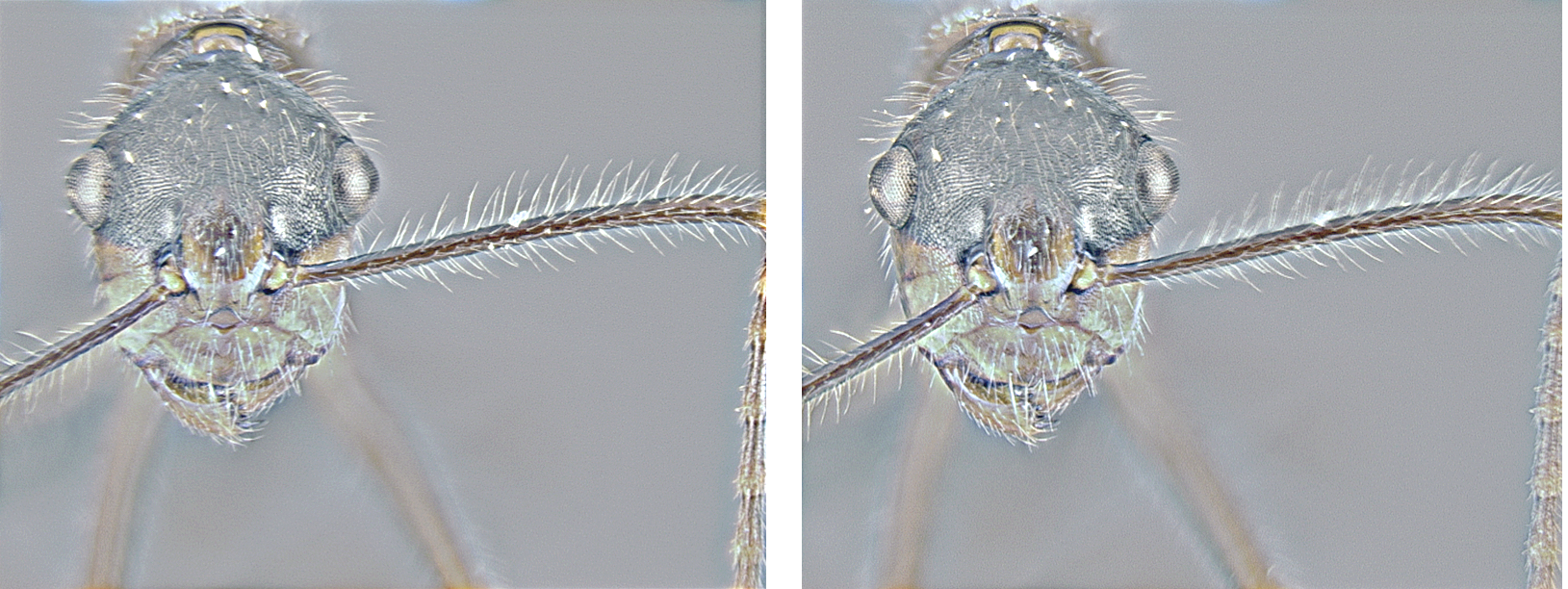}
    \caption{Stereo image of the ant. Ready for cross eye viewing.}
    \label{fig: ant_left_right}
\end{figure}

We implement the viewer using the Van Gogh framework by Adobe Systems Inc. Van Gogh is a high performance cross-platform application framework, focused on image processing and 2D/3D rendering. We place the images in an array, at equal distances from each other. Because we use orthographic projection, all images are of the same size regardless of their position. The virtual camera is placed in front of the array. From front to the back, we set the image alpha value to be $1/n, 1/(n-1), 1/(n-2), ..., 1$ where $n=13$. During alpha blending each of the images will contribute effectively $1/13$ of the final rendering result. A left/right comparison is shown in Fig.\ref{fig: ant_left_right}.

We can view parallax of the scene by moving the camera around. The change of camera position is equivalent to appropriate shifts of the focal stack images. The $3D$ effect is artifact free within a certain solid angle around the center. In the absence of occlusions, we believe larger viewing angles, up to $2\pi$, can be achieved with larger number of layers positioned at higher density. 

\section{The John Equation}

One can verify by differentiation that the radiance $r$ in Eq.\ref{eqn: John Transform} must satisfy John's equation \cite{john1985ultrahyperbolic}:
\begin{align}
    (\frac{\partial}{\partial y} \frac{\partial}{\partial u} - \frac{\partial}{\partial x} \frac{\partial}{\partial v}) r = 0
    \label{eqn: John's equation}
\end{align}

\subsection{The Ultrahyperbolic Equation}

Furthermore, if we do the following reparametrization of $r(x,y,u,v)$ into $R(\xi_1,\xi_2,\xi_3,\xi_4)$ (i.e. transform from $4D$ $(x,y,u,v)$-space to $4D$ $(\xi_1,\xi_2,\xi_3,\xi_4)$-space),
\begin{align}
    \begin{cases}
    \xi_1 = \frac{1}{2}(u + y)\\
    \xi_2 = \frac{1}{2}(u - y)\\
    \xi_3 = \frac{1}{2}(v + x)\\
    \xi_4 = \frac{1}{2}(v - x)
    \end{cases}
    \qquad \qquad
    \begin{cases}
    x = \xi_3 - \xi_4\\
    y = \xi_1 - \xi_2\\
    u = \xi_1 + \xi_2\\
    v = \xi_3 + \xi_4
    \end{cases}
    \label{eqn: linear transform}
\end{align}
we will get the ultrahyperbolic partial differential equation
\begin{align}
    (\partial_{\xi_1\xi_1} - \partial_{\xi_2\xi_2} - \partial_{\xi_3 \xi_3} + \partial_{\xi_4 \xi_4})R =0 \qquad \text{ or equivalently, } (\Delta_{14} - \Delta_{23}) R = 0,
    \label{eqn: ultrahyperbolic}
\end{align}
where $\Delta_{14}$ and $\Delta_{23}$ are the Laplacians in the $(\xi_1,\xi_4)$ and $(\xi_2,\xi_3)$ planes, respectively.

\subsection{The Dimensionality Gap}
If we perform Fourier transform $\mathcal{F}$ in Eq.\ref{eqn: John's equation} and Eq.\ref{eqn: ultrahyperbolic}, 
\begin{align*}
    \mathcal{F}: r(x,y,u,v) &\longrightarrow \tilde{r}(k_x,k_y,k_u,k_v)\\
    \mathcal{F}: R(\xi_1,\xi_2,\xi_3,\xi_4) &\longrightarrow \tilde{R}(k_{\xi_1}, k_{\xi_2}, k_{\xi_3}, k_{\xi_4})\\
    \partial_y \partial_u r - \partial_x \partial_v r = 0 &\longrightarrow (k_y k_u - k_x k_v)\tilde{r} = 0 \\
    (\partial_{\xi_1}\partial_{\xi_1} - \partial_{\xi_2}\partial_{\xi_2} - \partial_{\xi_3}\partial_{\xi_3} + \partial_{\xi_4}\partial_{\xi_4})R=0 &\longrightarrow (k_{\xi_1}^2 - k_{\xi_2}^2 - k_{\xi_3}^2 + k_{\xi_4}^2) \tilde{R} = 0.
\end{align*}
We see that $\tilde{r} = 0$ and $\tilde{R} = 0$ everywhere except on a $3D$ submanifold inside $4D$ frequency space. 
That is the so called ``Dimensionality Gap'' of Ng \cite{Ng:2005:FSP:1073204.1073256} and Levin-Durand \cite{Levin:2009:FAC:1576246.1531403}.

The radiance in frequency domain is a generalized function, related to Dirac's $\delta$-function, having support on the above manifold of measure zero. The manifold is actually the ultrahyperbolic cone in $4D$. There are different representations of this cone as arrays of microimages in the frequency domain, but the practical results have been achieved only after explicitly taking it into consideration in related image rendering code \cite{Levin:2009:FAC:1576246.1531403, 5539854}.

\section{Asgeirsson's Theorems}
Next we use the ultrahyperbolic equation with four variables as shown in Eq.\ref{eqn: ultrahyperbolic}. We base our analysis on the following theorems by Asgeirsson (see \cite{courant2008methods}). 

\begin{theorem}
Integral over a circle $C_1$ with radius $R$ in the $(1,4)$-plane is equal to the integral over the same radius circle $C_2$ in the $(2,3)$-plane, i.e.,
\begin{align}
     \int_{0}^{2\pi} r(\xi_1 + R \cos \theta,\xi_2,\xi_3,\xi_4 + R \sin \theta) \dd \theta = \int_{0}^{2\pi} r(\xi_1, \xi_2 + R \cos \theta,\xi_3+ R \sin \theta, \xi_4) \dd \theta
\end{align}
\end{theorem}

\begin{theorem}
More generally, if we consider a double integral over two circles, one of which has radius $R_1$ in $(1,4)$-plane and the other has radius $R_2$ in $(2,3)$-plane, is equal to the double integral over two circles with two radii switched in the two planes, i.e.,
\begin{align*}
     &\int_0^{2\pi} \int_{0}^{2\pi} r(\xi_1 + R_1 \cos \theta_1,\xi_2 + R_2 \cos \theta_2,\xi_3 + R_2 \sin \theta_2,\xi_4 + R_1 \sin \theta_1) \dd \theta_1 \dd \theta_2 \\&= \int_{0}^{2\pi} \int_{0}^{2\pi} r(\xi_1 + R_2 \cos \theta_1,\xi_2 + R_1 \cos \theta_2,\xi_3 + R_1 \sin \theta_2,\xi_4 + R_2 \sin \theta_1) \dd \theta_1 \dd \theta_2
\end{align*}
\end{theorem}



Next we discretize John's equation Eq.\ref{eqn: John's equation} and the ultrahyperbolic equation Eq.\ref{eqn: ultrahyperbolic}, as well as the equations for the two Asgeirsson theorems, to produce the corresponding processing kernels.

\section{Kernels} \label{sec: kernel derivation}

\subsection{Laplacian}

\begin{figure}[hbt!]
    \centering
    \includegraphics[scale = 0.25]{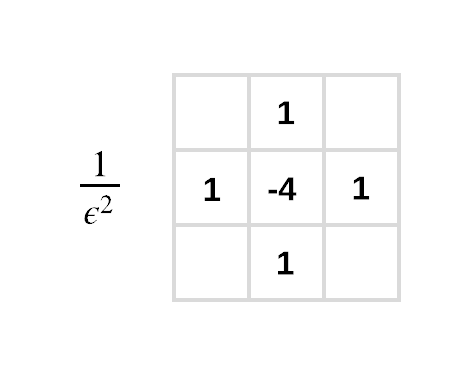} \qquad
    \includegraphics[scale = 0.25]{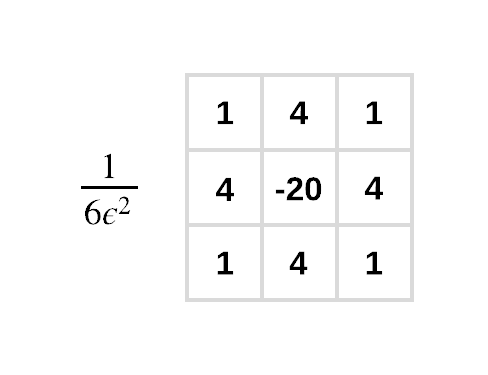}
    \caption{$3 \times 3$ Discrete Laplacian kernels ($\epsilon$ is the distance between adjacent cells): left is second order approximation of $\Delta$; right is also second order but with higher-order rotational invariant property, i.e., more stable and robust.}
    \label{fig:LaplacianKernel1}
\end{figure}

The 2D Laplace operator, i.e., $\Delta f = f_{xx} + f_{yy}$ in Euclidean space has rotational invariance. However, the discrete Laplacian may not have such nice property. The Taylor expansion  results\cite{horn1986robot} of Fig.\ref{fig:LaplacianKernel1}:
\begin{align*}
    \text{Left of Fig.\ref{fig:LaplacianKernel1}} &= \Delta + \frac{\epsilon^2}{12} (\frac{\partial^4}{\partial x^4} + \frac{\partial^4}{\partial y^4}) + O(\epsilon^4)\\
    \text{Right of Fig.\ref{fig:LaplacianKernel1}} &= \Delta + \frac{\epsilon^2}{12} \Delta(\Delta) + O(\epsilon^4)
\end{align*}
The differential operator $\frac{\partial^4}{\partial x^4} + \frac{\partial^4}{\partial y^4}$ does not have the rotational invariance, while $\Delta(\Delta)$ does. Therefore, the right kernel in Fig.\ref{fig:LaplacianKernel1} is more stable and robust in terms of rotation.

\subsection{Kernels for John's Equation}
 
\begin{figure}[hbt!]
    \centering
    \includegraphics[scale = 0.15]{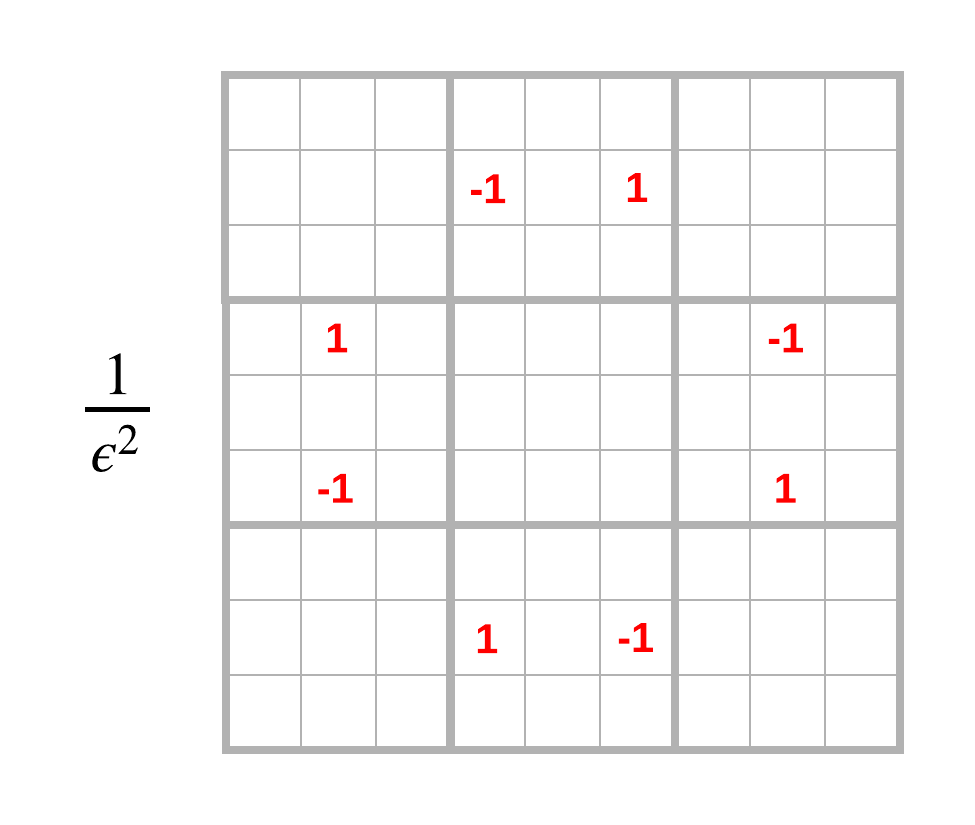} \qquad
    \includegraphics[scale = 0.15]{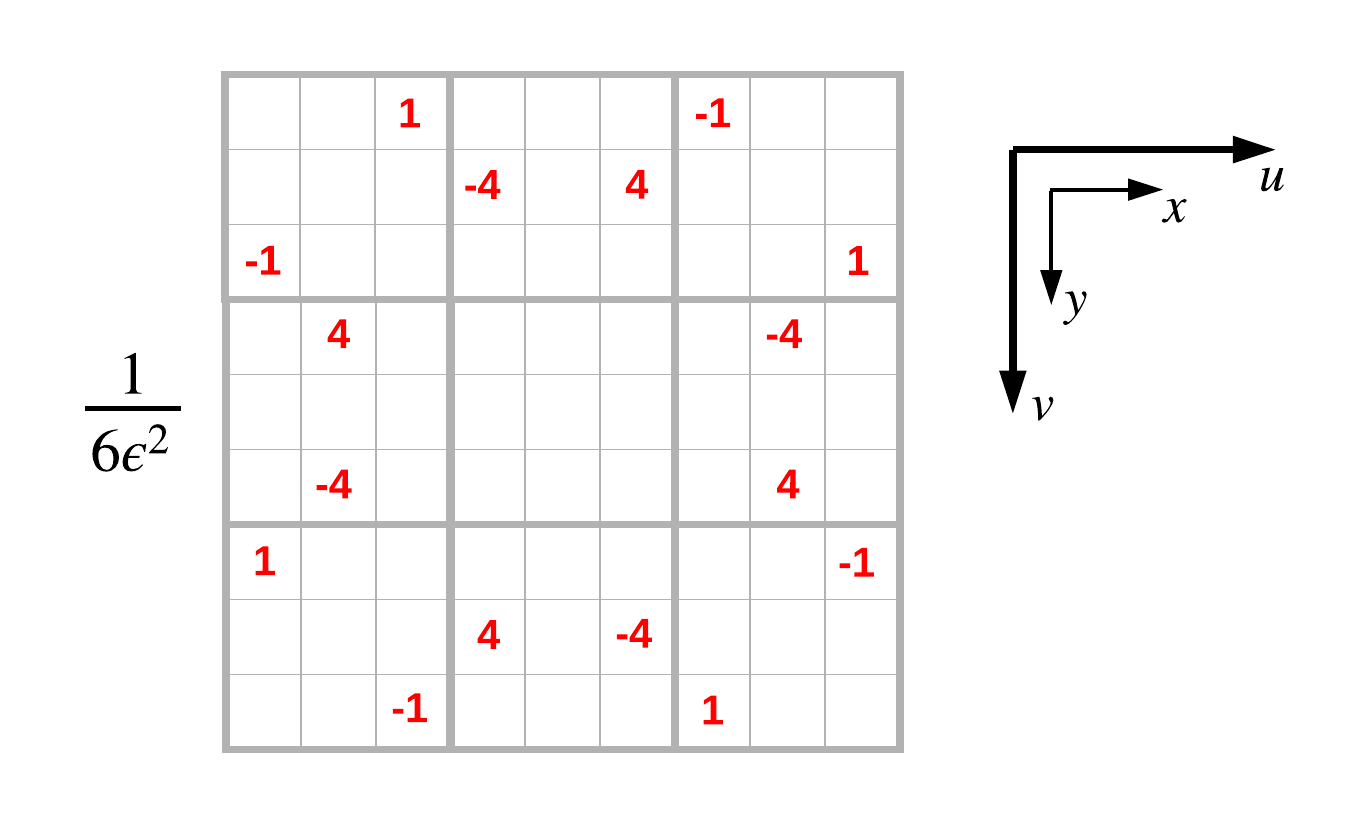}
    \caption{$9 \times 9$ Discrete John's kernels in 4D discrete grids (obtained by assigning 2D array within another 2D array): Based on different Laplacian kernels in Fig.\ref{fig:LaplacianKernel1}.}
    \label{fig:John's kernel}
\end{figure}

By combining two identical Laplacian kernels, we can get the kernel of ultrahyperbolic equation and we have two choices in Fig.\ref{fig:LaplacianKernel1}. 
In order to get the kernel of John's equation, we simply perform the linear transformation, as defined in Eq.\ref{eqn: linear transform}, to convert everything back to the lightfield coordinate $(x,y,u,v)$. The resulted kernels are  shown in Fig.\ref{fig:John's kernel}.

\subsection{Asgeirsson's perspective}
\begin{figure}[hbt!]
    \centering
    \includegraphics[scale = 0.2]{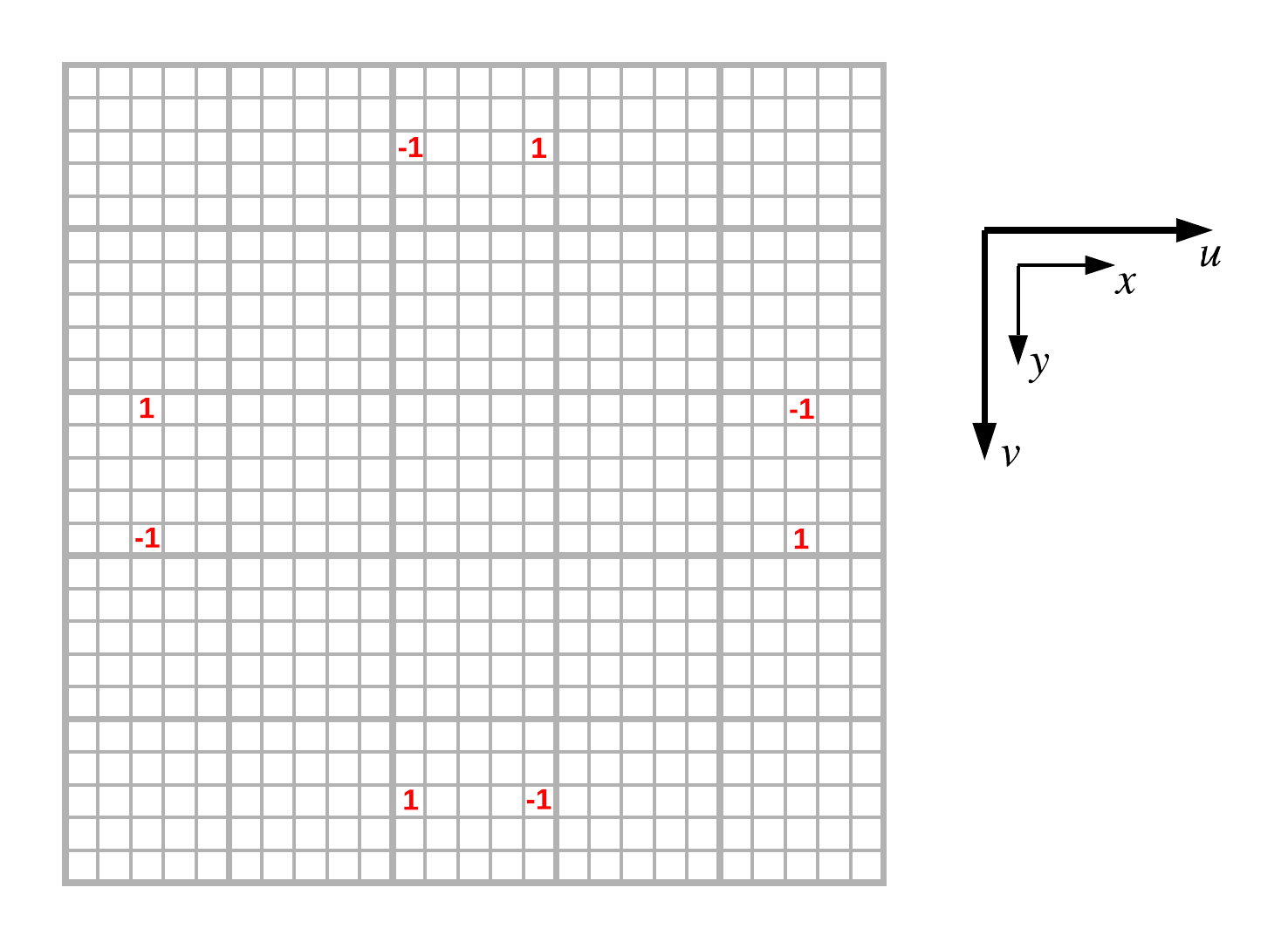}
    \caption{Kernel of Asgeirsson Theorem 1.1 with radius $R = 2$.}
    \label{fig: AsgR2}
\end{figure}

By the Asgeirsson's theorem and Eq.(5), we know the integration of $f$ over the unit circle in $(\xi_1,\xi_4)$-plane should be the same with the integration of $f$ over the unit circle in $(\xi_2,\xi_3)$-plane.
The parameterization of two unit circles in $(u,v,x,y)$-space can be represented by,
\begin{align}
    c_1: 
    \begin{cases}
    x = -\sin \varphi \\
    y = \cos \varphi \\
    u = \cos \varphi \\
    v = \sin \varphi 
    \end{cases}
    \qquad\qquad
    c_2: 
    \begin{cases}
    x = \sin \varphi \\
    y = -\cos \varphi \\
    u = \cos \varphi \\
    v = \sin \varphi 
    \end{cases}
    \label{eqn: two unit circles}
\end{align}
If we use four integer points to sample the circles, i.e., $\varphi = 0, \frac{\pi}{2}, \pi, \frac{3}{2}\pi$ in Eq.\ref{eqn: two unit circles}, we can also attain the John's kernel, as shown in the left of Fig.\ref{fig:John's kernel}. Therefore, John's kernel also verifies Theorem 1.1 with radius $1$. See Fig.\ref{fig: AsgR2} for the case of radius $2$.

\section{Characteristics of the Ultrahyperbolic Differential Equation}

The characteristic solution of John's equation (Eq.\ref{eqn: John's equation}) is 
\begin{align}
    r(x,y,u,v) = g(x - cu, y - cv), \qquad \forall  c \in \R
\end{align}
It is easy to check that any function $g(x-c u, y-c v) \in C^2(\R^2)$ satisfies Eq.\ref{eqn: John's equation}. This is a large family of solutions. It is representative of the repetitive nature of lightfields captured as 2D arrays of microimages where $u$ and $v$ enumerate microimages. Features are ``shifted'' or ``travel'' from microimage to microimage; different features travel at different ``speed''. This is the most visible fingerprint of lightfields. Notice that the speed of such ``ultrahyperbolic waves'' $c$ can be any real number. Compare to the wave equation which describes waves with one fixed speed $c$ which is a parameter in the wave equation. In John's equation there is no parameter $c$. This behavior is clearly observed in our experiments (see Fig.\ref{fig: LightfieldCoordinate} as an example).

If we change the coordinates by Eq.\ref{eqn: linear transform}, i.e., $(x,y,u,v) \longrightarrow (\xi_1,\xi_2,\xi_3,\xi_4)$, we have the characteristic solution of the ultrahyperbolic equation:
$$f(\xi_1,\xi_2,\xi_3,\xi_4) = g(-c\xi_1 - c\xi_2  + \xi_3 - \xi_4, \xi_1-\xi_2-c\xi_3 -c\xi_4)$$ 



\section{Lightfield Experiments}

\subsection{Lightfield Viewer}

\begin{figure}[hbt!]
    \centering
    \includegraphics[scale = 0.5]{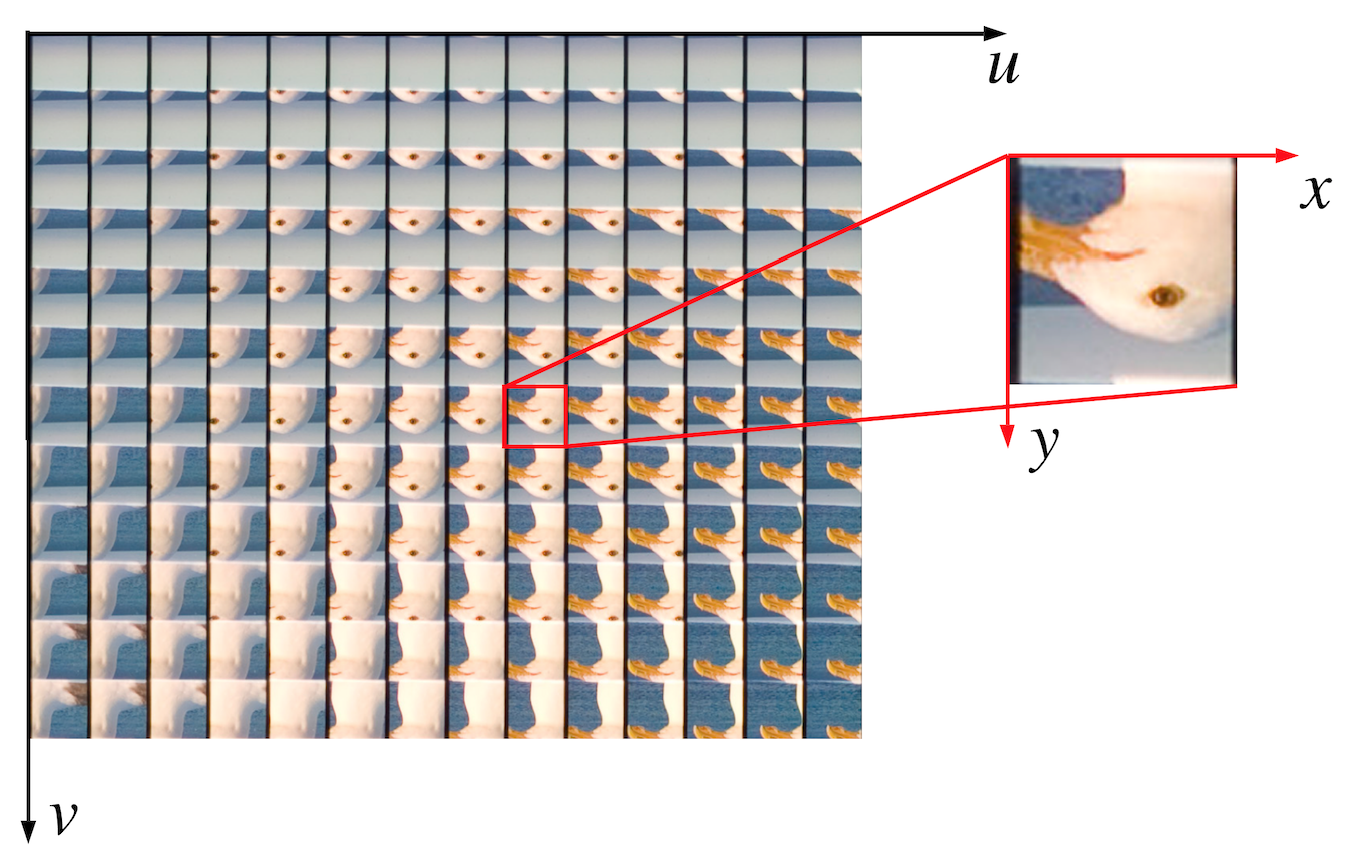}
    \caption{Example Lightfield and Coordinates}
    \label{fig: LightfieldCoordinate}
\end{figure}

The lightfield image used in our experiment is a seagull scene of $72\times96$ microimages with each of size $74.77\times74.77$ pixel. The coordinates are shown in Fig.\ref{fig: LightfieldCoordinate}.

A lightfield viewer is to render the lightfield image under some desirable parameters. The major job is to implement correct sampling. A raw lightfield image consists of many micro-images. The neighboring micro-image features are shifted by a small amount of parallax. Therefore, same objects have certain offsets in different neighboring micro-images. This is in line with our family of characteristic solutions as described in section 6. 

\begin{figure}[hbt!]
    \centering
    \includegraphics[scale = 0.45]{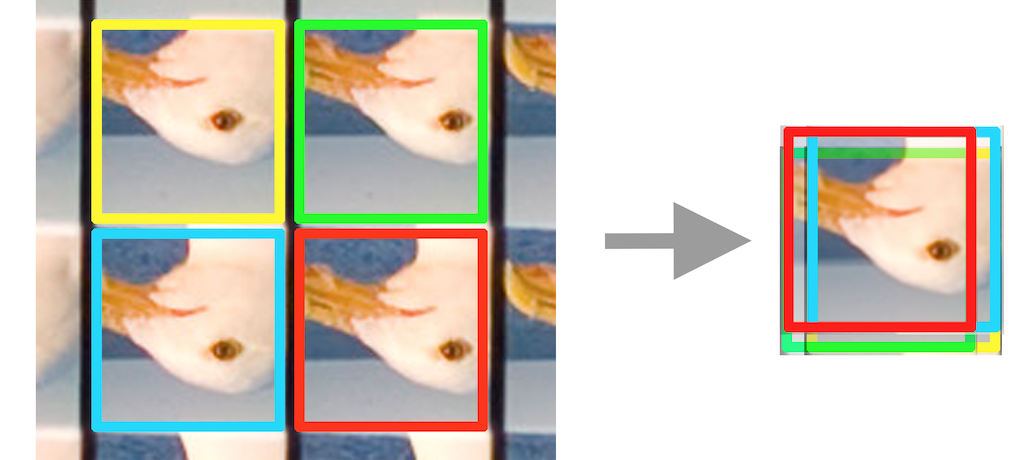}
    \caption{Lightfield Blending}
    \label{fig: LightFieldBlend}
\end{figure}

In order to match image content, one needs to shift the micro-images appropriately. Note that for foreground and background, shifts are different. After appropriate shifting, one can blend the micro-images to get the final rendered result as shown in Fig.\ref{fig: LightFieldBlend}.

\subsection{John's Kernel and Depth Estimation}

Section \ref{sec: kernel derivation} derives John's kernel from Asgeirsson's theorem. If John's equation is satisfied, then the resulting image after applying John's kernel should be black. However, on a real captured lightfield the microimage content is not directly corresponding to each other at the subpixel level. That is due to the big and different offset of foreground and background. The offset needs to be on a subpixel level and actually needs to satisfy the Nyquist condition, while in reality it is chosen (by the camera design) to be several pixels. In other words, the captured lightfield is sparsely sampled, and severely aliased in the angular dimensions $u$ and $v$ in order to maximize spatial resolution. (It is like trying to compute fine level gradient in an image which is reduced in size with ``nearest neighbor'' sampling where only one out of 10 pixels is left and information about smooth changes is lost.)

Thus, our kernels must be applied to properly shifted microimages that compensate for the above aliasing. The shift value is $F+D$ as given in previous section. Because each time only those in focus subsets will satisfy the kernel, we set the kernel shift to use the blend shift, which is just set $D=0$. In other word, only the in focus part will be black.

The method above yielded a new lightfield image transformed by John's kernel. Then we blended this new image. When changing the focus(kernel shift), it looks like a ``shadow'' is coming through the scene. 
The position of the shadow indicates that this region is currently in focus. One direct intuition is to estimate depth by using the shadow. 

\begin{figure}[hbt!]
    \centering
    \includegraphics[scale = 0.85]{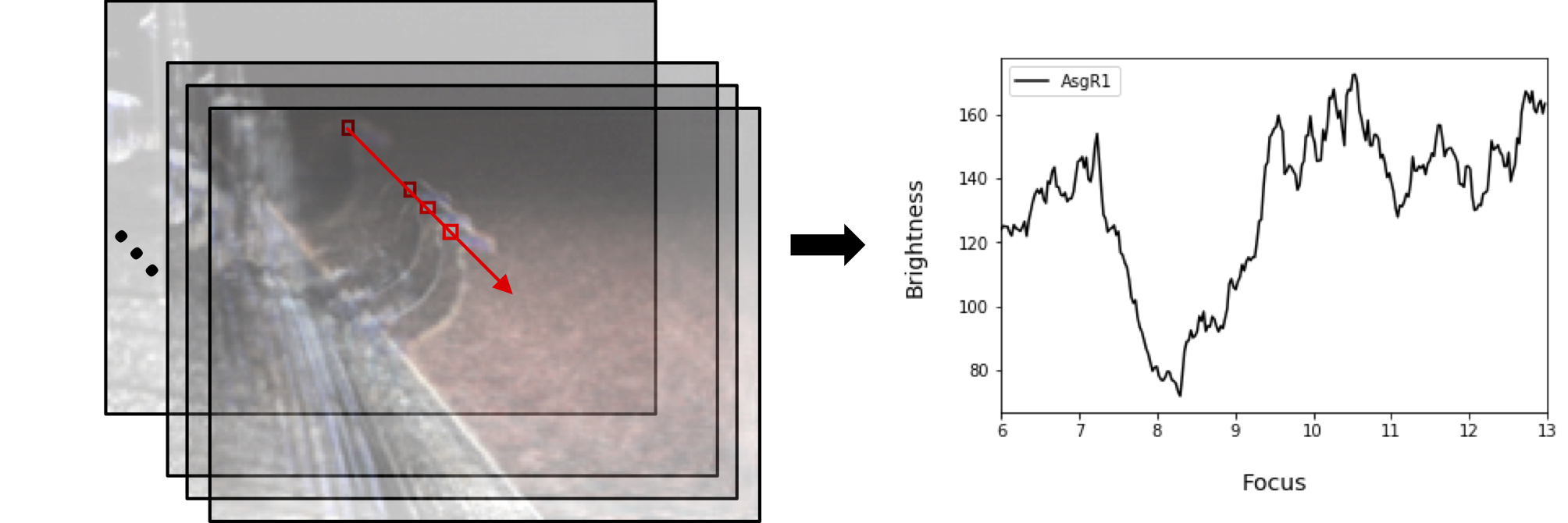}
    \caption{Brightness Focus Relationship. For each pixel on transformed blended image stack, the brightness changes according to focus.}
    \label{fig: brightness_focus_pixel_stack}
\end{figure}

We change the $F$ from $6$ to $13$ and output a stack of 256 blended images. The output images size are normalized to $720 \times 540$, so we can correspond pixels among images. For each pixel, we can get brightness among different focus image, as shown in Fig.\ref{fig: brightness_focus_pixel_stack}.

\begin{figure}[hbt!]
    \centering
    \includegraphics[width = \textwidth]{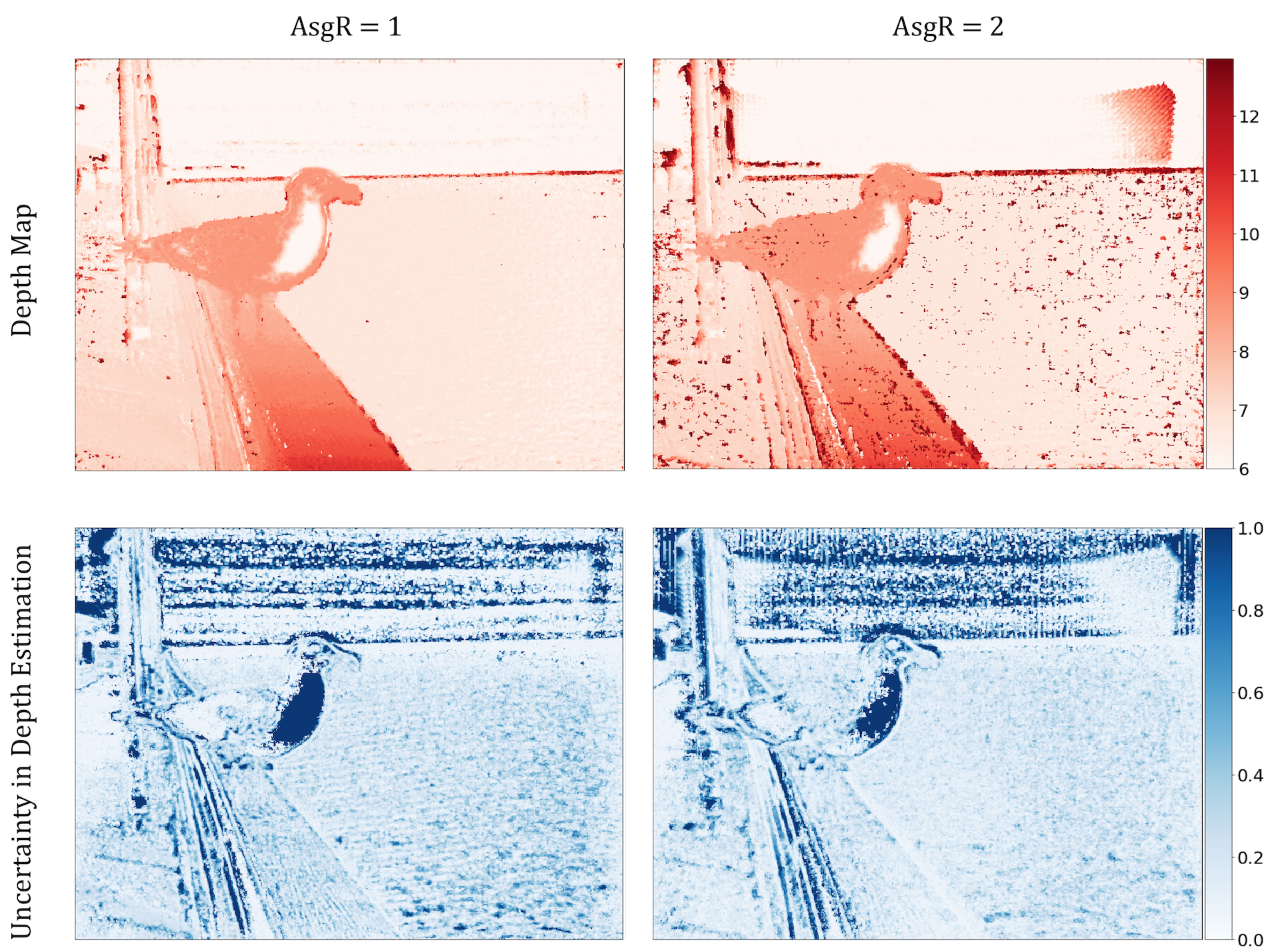}
    \caption{Depth and Uncertainty in Depth Estimation for AsgR1 and AsgR2.
    For depth map, AsgR1 on the left shows less noise. AsgR1 generally has larger uncertainty in depth.}
    \label{fig: asgfocusmap}
\end{figure}

After loading the 256 images, we compute the depth map by comparing and writing the focus of the smallest brightness image to the map. Replicating this procedure for all pixels on images we can get the depth maps for both kernels of Asgeirsson $R=1$ and $R=2$. As shown in Fig.\ref{fig: asgfocusmap}. The right two plots for $R=2$ has ``larger'' seagull than the $R=1$ on left. This is because $R=2$ discards outer 2 layers of microimages while $R=1$ discards 1 layer, which results in the size difference.

\begin{figure}[hbt!]
    \centering
    \includegraphics[width = \textwidth]{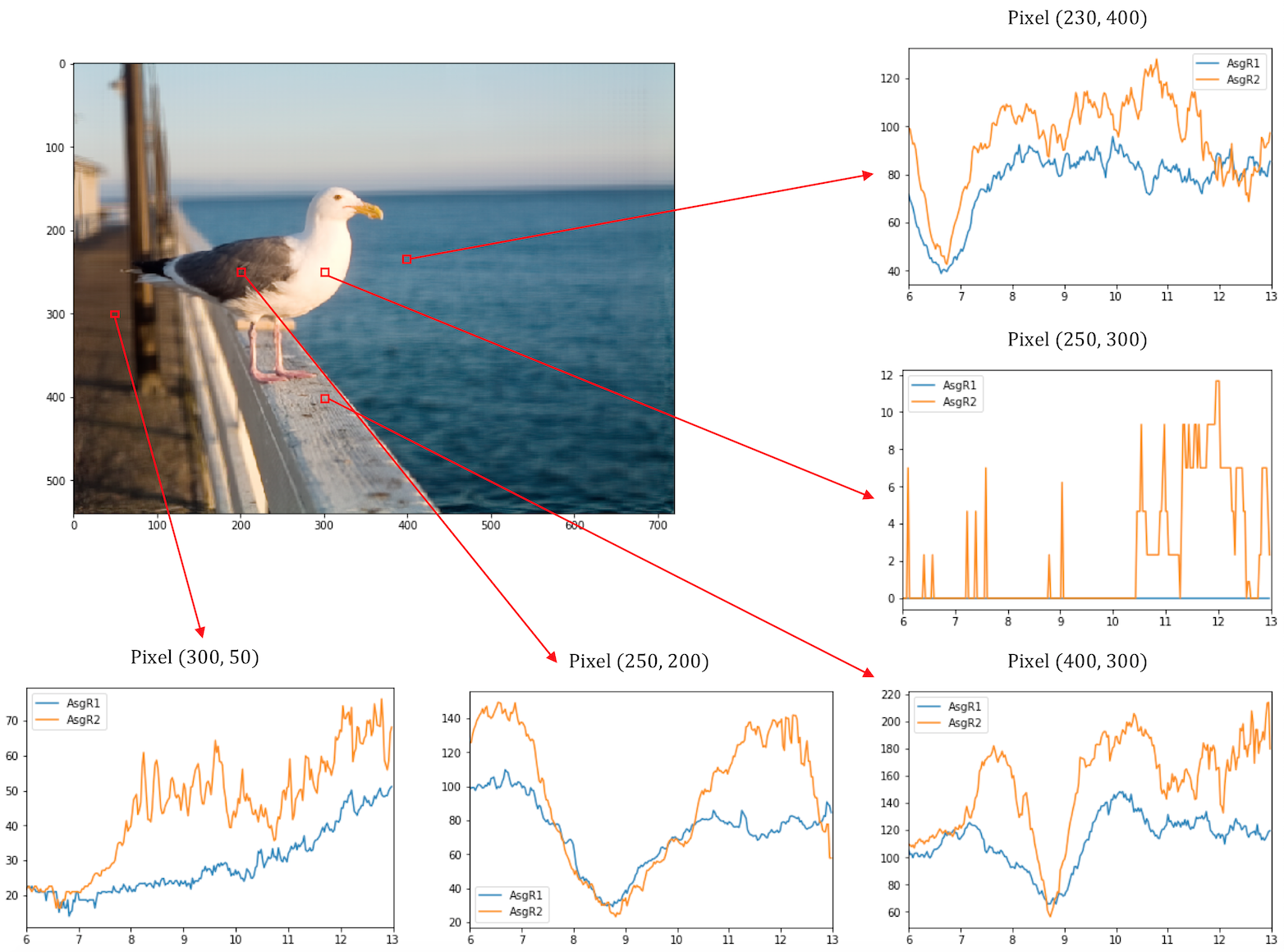}
    \caption{Pixel Brightness Stack - Depth relationship. The figure was obtained under $B=45$. Titles for each subplot are pixel position on stacked image. Blue and orange lines are for AsgR1 and AsgR2. Horizontal and vertical are for depth and brightness. }
    \label{fig: pixel_stack}
\end{figure}

One interesting observation is that the duration or ``size'' of the shadow is not uniform across one image. Also, for same position on different image of $R=1$ and $R=2$, duration also varies. The duration is important because it indicates uncertainty in depth estimation. In order to qualitatively visualize the effect, the pixel brightness and depth relationship are plot for 5 selected position in Fig.\ref{fig: pixel_stack}. For orange lines of AsgR2, they have narrow troughs on beam and sea. On the seagull chest, the orange line have peculiar behavior because the image is of approximately constant colors. In other words, the image lacks depth information. For far away object like the pavement, brightness increases monotonically as the depth is beyond the focus range.

To get more quantitative result of the uncertainty, we adopt a simple algorithm of measuring the width. For normal case, we start from the minimal brightness focus and expand the depth. We record the first two depths, where the brightness is $L_{min}+0.05*(L_{max}-L_{min})$. The depth range of these two depths are used for the uncertainty. Two special cases are considered here. Firstly, when the depth range is large than 1, we clipped the value to 1 for plotting. Secondly, for region with same color, the brightness has multiple zeros. We add up the range for all zeros to get the uncertainty. The results were plotted in Fig.\ref{fig: asgfocusmap}. From Fig.\ref{fig: asgfocusmap}, the AsgR1 has much darken blue on the sea and the beam, which indicates that it has larger uncertainty on these regions. Therefore, the above experiments states the John's equation is satisfied on lightfield image. Overall, the AsgR2 has smaller depth uncertainty than AsgR1, which means the depth estimation is of better quality.

\section{Conclusions}
In $3D$ space there are 2 types of Radon transform – based on lines or planes. We show that the line transform, also known as John's transform, actually describes the lightfield. We also show that by applying the inverse John transform to a focal stack captured by a conventional camera we can generate virtual $3D$ images and view them from a wide range of angles. This generalizes effects similar to previous results known under the name ``Dimensionality Gap'' and puts the problem into the much older and well studied framework of Radon transforms and applications.

From another perspective, we show that John’s transform satisfies a PDE knows as John's equation. This equation is the constraint that produces the dimensionality gap. We observe that it also describes the repetitive structure of captured lightfields. This equation explains why lightfields look like a recording of ``running waves'' with features moving away from each other. Unlike what is observed in the wave equation, John's equation permits any speed and multiple speeds of different features.

John’s equation can be used to generate a range of filters, for example filters that compute depth from lightfield. This type of result is specific to the $4D$ lightfield, and not equivalent to other methods of computing depth. This is possible because John’s equation is equivalent to the ultrahyperbolic equation, for which there are several global theorems. We show how to use these theorems to generate the above sequence of kernels, of different sizes, which kernels can be applied as lightfield filters. Our results are independent of previously known filters in the lightfield, and can have different applications. We only provide the example of how they can be used to compute depth at better quality, and show that bigger filters produce better results.

\newpage
\section{Appendix}



\begin{figure}[hbt!]
    \centering
    \includegraphics[scale = 0.15]{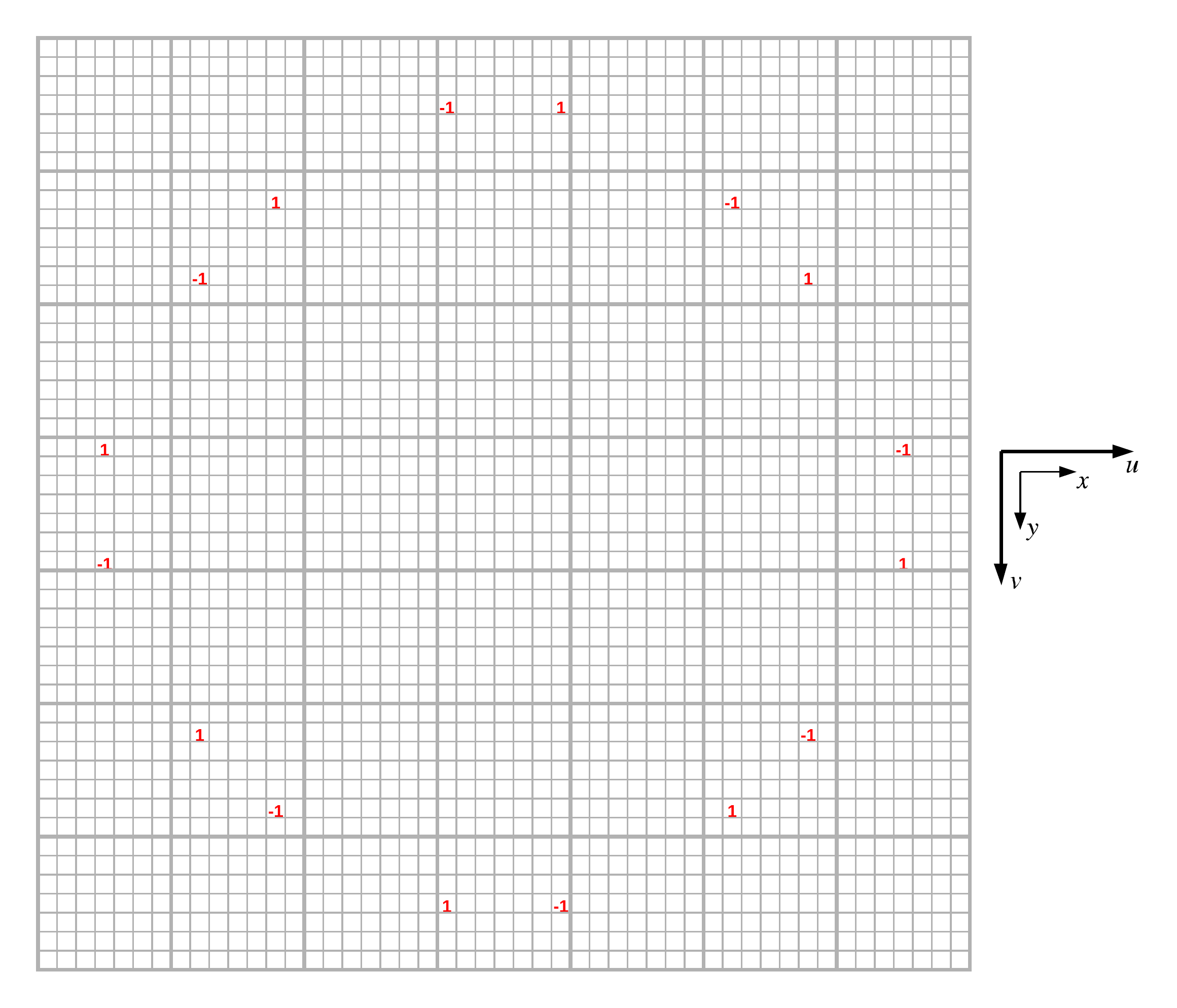}
    \caption{Kernel of Asgeirsson's Theorem 1.1 with radius $R = 3$.}
    \label{fig: AsgR3}
\end{figure}

\begin{figure}[hbt!]
    \centering
    \includegraphics[scale = 0.15]{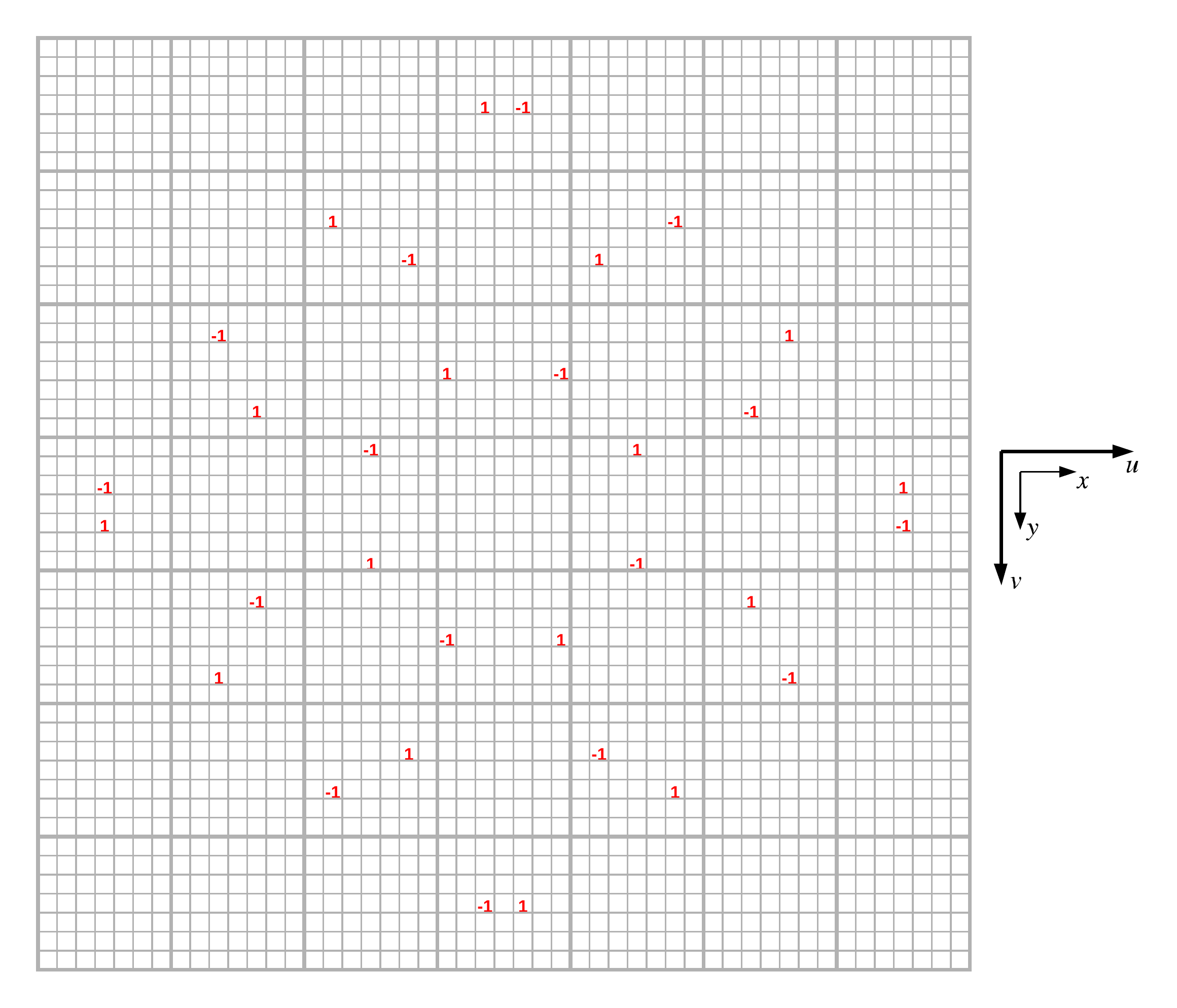}
    \caption{Kernel of Asgeirsson's Theorem 1.2 with radius $R_1 = 1, R_2 = 2$.}
    \label{fig: AsgT2R1R2}
\end{figure}


\clearpage
\bibliographystyle{siam}
\bibliography{ref}

\end{document}